\documentclass[10pt]{article}
\usepackage{amsmath,amssymb,graphicx,slashbox}
\usepackage{color}

\def\mydate{March 17, 2010}
\def\ignore#1{{}}

\newcommand{\beeq}{\begin{equation}}
\newcommand{\eneq}{\end{equation}}
\newcommand{\beqn}{\begin{eqnarray}}
\newcommand{\eeqn}{\end{eqnarray}}

\def\mybig{\displaystyle \strut }

\def\dd{\partial}
\def\la{\raise.16ex\hbox{$\langle$}\lower.16ex\hbox{}  }
\def\ra{\raise.16ex\hbox{$\rangle$}\lower.16ex\hbox{} }
\def\go{\rightarrow}

\def\onehalf{ \hbox{${1\over 2}$} }

\def\eff{{\rm eff}}

\def\EM{{\rm EM}}

\def\KK{{\rm KK}}

\def\psibar{ \psi \kern-.65em\raise.6em\hbox{$-$} }
\def\psibarl{ \psi \kern-.65em\raise.6em\hbox{$-$} \lower.6em\hbox{} }

\def\myfrac#1#2{{\mybig #1\over \mybig #2}}

\def\mysection#1{{\vskip 25pt \leftline{\large \bf #1} \vskip 5pt }}

\begin{document}

{\small \noindent \mydate    \hfill OU-HET 656/2010}

\vskip 1.1cm

\begin{center}

\baselineskip=15pt plus 1pt minus 1pt

{\large \bf Gauge-Higgs Unification:\\
Stable Higgs Bosons as Cold Dark Matter\footnote{To appear in the Proceedings of
{\it ``SCGT09: Strong Coupling Gauge Theories in LHC Era'',} 
Nagoya, Japan, 8-11 December, 2009. }}

\vskip 10pt

{\small Yutaka Hosotani}

\baselineskip=13pt plus 1pt minus 1pt

{\small \it Department of Physics, Osaka University\\
Toyonaka, Osaka 560-0043, Japan}
\end{center}

\vskip 3pt

\begin{abstract}
In the gauge-Higgs unification the 4D Higgs field 
becomes a part of the extra-dimensional component of the gauge potentials.
%being unified with the 4D gauge fields.  
In the $SO(5) \times U(1)$ gauge-Higgs unification in the Randall-Sundrum
warped spacetime the electroweak symmetry is dynamically broken through 
the Hosotani mechanism.  The Higgs bosons become absolutely stable,
and become the dark matter of the universe.
The mass of the Higgs boson is determined from the WMAP data to be
about 70 GeV.
\end{abstract}

\centerline{\small {\it Keywords}: Higgs boson, dark matter, gauge-Higgs unification, 
Hosotani mechanism.}

%\vskip .5cm 

\baselineskip=13pt plus 1pt minus 1pt

\mysection{1. Introduction}

%\section{Introduction}
Does the Higgs boson exist?  What is it really like?
%What is the Higgs boson? Does it really exist?  
What constitutes the dark matter in the universe?
These are two of the most important problems in current physics.  
We would like to point out that these two mystery particles
are really the same.  In the gauge-Higgs unification scenario
Higgs bosons become absolutely stable, and become the dark matter
of the universe. 

In the gauge-Higgs unification the 4D Higgs boson is identified with 
a part of the extra-dimensional component of the gauge potentials.
Its couplings with others particles are controlled by the gauge
principle.   The 4D Higgs field corresponds to 4D fluctuations of
an AB phase in the extra dimensions.\cite{YH1, YH2, Davies1}  
In the $SO(5) \times U(1)$ gauge-Higgs
unification model in the Randall-Sundrum warped space 
the AB phase $\theta_H$  takes exactly the value $\onehalf \pi$ in the vacuum 
as a consequence of  quantum dynamics.  At this particular value of $\theta_H$
the 4D Higgs boson becomes absolutely stable.  

The relic abundance of the Higgs bosons in the present universe
is evaluated definitively with the mass of the Higgs boson as
the only relevant variable parameter.  Astonishingly the average mass density of the dark 
matter determined  from the WMAP data is obtained with the Higgs mass 
around 70~GeV.  It does not contradict with the LEP2 bound, because
the $ZZH$ coupling vanishes at $\theta_H = \onehalf \pi$.  
The gauge-Higgs unification scenario gives 
a completely new viewpoint for the Higgs boson.\cite{HKT}
Further it gives definitive predictions for electroweak gauge couplings
of quarks and leptons, which can be tested experimentally.\cite{HNU}

\mysection{2. $SO(5) \times U(1)$ gauge-Higgs unification in RS}

%\section{$SO(5) \times U(1)$ gauge-Higgs unification in RS}
 
 We consider an $SO(5) \times U(1)$ gauge theory in the five-dimensional
 Randall-Sundrum (RS) warped spacetime.\cite{HKT}-\cite{HK}
 Its metric is given by 
 \beeq
 ds^2 
  = e^{-2\sigma(y)}\eta_{\mu\nu} dx^\mu dx^\nu + dy^2 , \qquad
 \label{metric1}
\eneq
where $\eta_{\mu\nu} =\textrm{diag}(-1,1,1,1)$,
$\sigma(y)=\sigma(y+2L)$, and $\sigma(y)=k|y|$ for $|y|\leq L$.
The fundamental region in the fifth dimension is given by $0\leq y\leq L$.
The Planck brane and the TeV brane are located at $y=0$  and $y=L$, respectively.  
The bulk region $0 < y < L$ is  an anti-de Sitter spacetime with a 
cosmological constant  $\Lambda = - 6k^2$.

The RS spacetime has the same topology as the orbifold $M^4 \times (S^1/Z_2)$.
Vector potentials $A_M(x,y)$ of the gauge group $SO(5)$ and $B_M(x,y)$ of
$U(1)$ satisfy the orbifold boundary conditions 
\beqn
&&\hskip -1cm 
\begin{pmatrix}    A_\mu \cr   A_y \end{pmatrix}  (x,y_j-y)
= P_j \begin{pmatrix}    A_\mu \cr   - A_y \end{pmatrix}   (x,y_j+y) P_j^{-1}  ~, \cr
\noalign{\kern 5pt}
&&\hskip -1cm 
\begin{pmatrix}    B_\mu \cr   B_y \end{pmatrix}  (x,y_j-y)
= \begin{pmatrix}    B_\mu \cr   - B_y \end{pmatrix}   (x,y_j+y)   ~, \cr
\noalign{\kern 5pt}
&&\hskip -.7cm 
P_j =\textrm{diag}(-1,-1,-1,-1,+1) ~,   ~~ (j=0,1),
\label{BC1}
\eeqn
where $y_0=0$ and $y_1=L$.  By the boundary conditions
the $SO(5)\times U(1)$ symmetry is reduced to 
$SO(4)\times U(1) \simeq   SU(2)_L \times SU(2)_R \times U(1)$.
The symmetry $SU(2)_R \times U(1)$ is further spontaneously broken 
by a scalar field $\Phi(x)$ on the Planck brane to $U(1)_Y$.
$\Phi(x)$ belongs to $(0,{1\over 2})$ representation
of $SU(2)_L \times SU(2)_R$.
The pattern of the symmetry reduction\cite{HS2, HOOS} 
is depicted in fig.\ 1.

\begin{figure}[b]
\begin{center}
\vskip -.5cm 
\includegraphics[height=5.5cm]{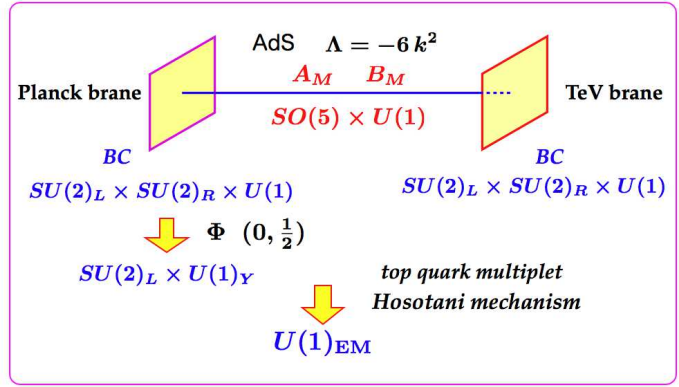}
\vskip -.4cm
\caption{The  symmetry reduction in the $SO(5) \times U(1)$ 
gauge-Higgs unification in RS. 
\label{symmetry}
}
\end{center}
\end{figure}

\begin{figure}[thb]
\begin{center}
%\vskip -.5cm
\includegraphics[height=4.cm]{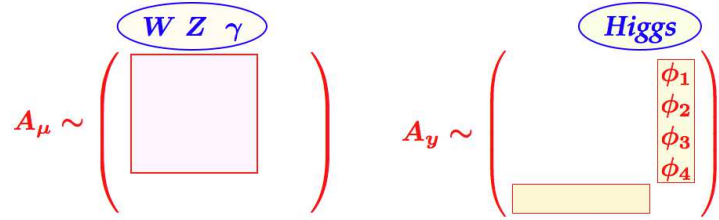}
\vskip -.5cm
\caption{$W,Z,\gamma$ appear as zero modes of $A_\mu$ whereas
the 4D Higgs field appears as  a zero mode of $A_y$.
\label{zero}}
\end{center}
\end{figure}

With the above boundary conditions the zero modes 
of 4D gauge fields reside in the 4-by-4 matrix part of $A_\mu$, 
whereas the zero modes of $A_y$  in the off-diagonal part 
of $A_y$.  The latter is an $SO(4)$ vector, or an $SU(2)_L$ doublet, 
corresponding to the Higgs doublet in the standard model. See fig.\ \ref{zero}.

\eject

Bulk fermions for quarks and leptons are introduced as multiplets in 
the vectorial representation of $SO(5)$.\cite{HNU, HK}
In the quark sector two vector multiplets are introduced for each 
generation.  In the lepton sector it suffices to introduce one multiplet
for each generation to describe massless neutrinos, whereas it is
necessary to have two multiplets to describe massive neutrinos.
Each vector multiplet $\Psi_a$ is decomposed to 
$(\onehalf, \onehalf ) \oplus (0,0)$ in the $SU(2)_L \times SU(2)_R$.
They satisfy the orbifold boundary condition
\beeq
\Psi_a  (x,y_j-y)
= P_j  \Gamma^5 \Psi_a  (x,y_j+y)  ~, 
\label{BC2}
\eneq
which gives rise to chiral fermions as zero modes.
The left-handed components of $(\onehalf, \onehalf )$ and the right-handed
components of $(0,0)$ have zero modes.

\begin{figure}[htb]
\begin{center}
\vskip -.5cm 
\includegraphics[height=9.cm]{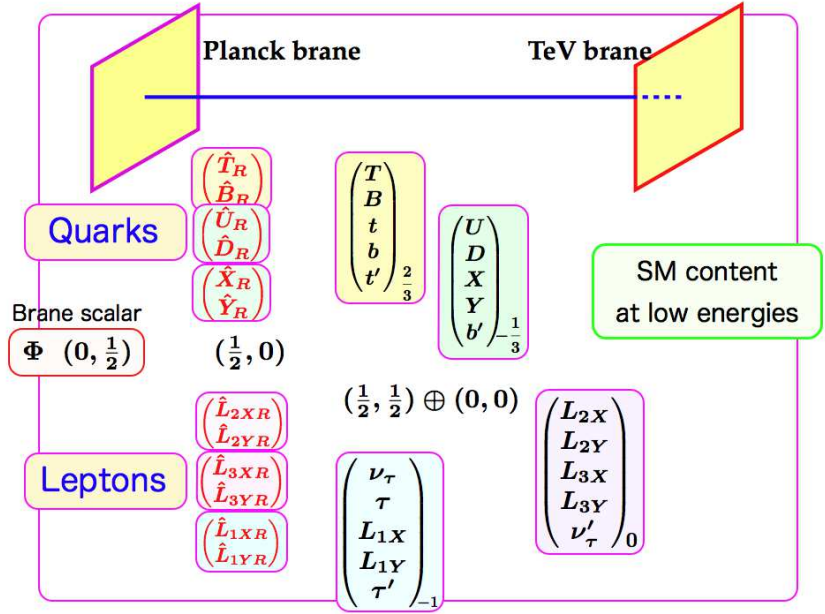}
\vskip -.5cm
\caption{Matter content in the model.
\label{matter}}
\end{center}
\end{figure}

In addition to the bulk fermions, right-handed brane fermions $\hat \chi_\alpha $ 
are  introduced on the Planck brane.  
The brane fermions $\hat \chi_\alpha $ belong to the $(\onehalf, 0)$ representation
of $SU(2)_L \times SU(2)_R$.
These brane fermions are necessary both
to have the realistic quark-lepton spectrum at low energies and to have the
cancellation of 4D chiral anomalies associated to the $SO(4) \times U(1)$ gauge
fields.     The matter content is summarized in fig.\ \ref{matter}.
The $SO(4) \times U(1)$ gauge invariance is maintained  on the Planck
brane as well.   The bulk fermions $\Psi_a$, the brane fermions $\hat \chi_\alpha$, 
and the brane scalar $\Phi$ form $SO(4) \times U(1)$ invariant interactions.
When $\Phi$ develops a non-vanishing expectation value to spontaneously 
break $SU(2)_R \times U(1)$ to $U(1)_Y$, it simultaneously 
gives mass couplings between $\Psi_a$ and $\hat \chi_\alpha$ 
which, in turn,  make all exotic fermions heavy.

\mysection{3. Higgs boson as an AB phase}

As shown in fig.\ \ref{zero}, the 4D Higgs field appears as a zero mode of $A_y$.
Without loss of generality one can suppose that the $A_y^{45}$ component 
develops a nonvanishing expectation value.
We write
\beeq
A_y (x,y) = \hat \theta_H (x) \cdot
 \sqrt{ \frac{4k}{z_L^2 -1} } ~ h_0(y)  \cdot
T^{\hat 4} + \cdots ~, ~~ z_L = e^{kL} 
\label{Ay1}
\eneq
where the zero-mode wave function is
$h_0(y) = [2k/(z_L^2 -1)]^{1/2} \, e^{2ky}$ ($0 \le y \le L$).
The generator $T^{\hat 4}$ is given by 
$(T^{\hat 4}_{\rm vec})_{ab} =( i /\sqrt{2}) 
 (\delta_{a5} \delta_{b4} -\delta_{a4} \delta_{b5}) $ in the vectorial representation
 and $T^{\hat 4}_{\rm sp} = ( 1 /2\sqrt{2}) I_2 \otimes \tau_1$ 
 in the spinorial representation.  
$\hat \theta_H(x)$ is decomposed as 
\beeq
\hat \theta_H (x) = \theta_H + \frac{H(x)}{f_H}
~~,~~
f_H = \frac{2}{g_A} \sqrt{\frac{k}{z_L^2 -1}}
\sim \frac{2}{\sqrt{kL}} \frac{m_\KK}{\pi g} ~~. 
\label{thetahat}
\eneq
Here the  Kaluza-Klein (KK) mass scale  is  
$m_\KK = \pi k z_L^{-1}$, and
the $SO(5)$ gauge coupling $g_A$ is related to the 4D $SU(2)_L$ weak 
coupling $g$ by $g=g_A/\sqrt{L}$.  

With $H(x) =0$, in the spinorial representation, 
\beeq
P \exp \bigg\{ ig_A \int^{L}_{0} dy   A_y  \bigg\}
= \exp  \bigg\{ \frac{i}{2} \theta_H  I_2 \otimes \tau_1  \bigg\}
\label{Wilson}
\eneq 
so that the constant part $\theta_H$ of $\hat  \theta_H(x)$ represents
an Aharonov-Bohm phase in the extra-dimension.
$\hat \theta_H$ is a phase variable.
There remains the residual gauge invariance
$A_M \go A_M' = \Omega A_M \Omega^{-1} - (i/g_A) \Omega \dd_M \Omega^{-1}$
which preserves the boundary conditions  (\ref{BC1}) and (\ref{BC2}).\cite{HM, HHHK}
In general,  new gauge potentials satisfy new boundary conditions
given by ${P_j'} = \Omega(x, y_j - y) P_j \Omega(x, y_j + y)^{-1}$. 
The residual gauge invariance is defined with $\Omega(x,y)$ satisfying
$P_j' = P_j$.  Consider a large gauge transformation
\beeq
\Omega^{\rm large}(y; \alpha) = \exp \bigg\{ - i\alpha \int_0^y dy \,
\sqrt{4k/(z_L^2-1)} ~ h_0(y) \cdot T^{\hat 4} \bigg\} ~.
\label{largeGT1}
\eneq
which shifts $\hat \theta_H (x)$ to
$\hat \theta_H' (x) = \hat \theta_H(x) + \alpha$.
With $\alpha = 2\pi$, $P_j' = P_j$ and 
$\hat \theta_H' = \hat \theta_H + 2\pi$.
 It implies that physics is invariant under
\beeq
\theta_H \go \theta_H + 2\pi ~~.
\label{periodicity}
\eneq

\mysection{4. Effective theory}

Four-dimensional fluctuations of the AB phase $\theta_H$ correspond to
the 4D neutral Higgs field  $H(x)$.   In the effective theory of the low energy
fields the Higgs field $H(x)$ enters always in the combination of 
$\hat \theta_H (x)$ in (\ref{thetahat}).   The effective Lagrangian 
must be invariant under $\hat \theta_H (x) \go \hat \theta_H (x) + 2\pi$.
The effective Higgs interactions with the 
$W$, $Z$ bosons,  quarks and leptons  at low energies are
summarized as \cite{HKT, HK, Sakamura1, Giudice1}
\beeq
{\cal L}_\eff  = - V_\eff (\hat \theta_H) 
                 - m_W^2(\hat \theta_H) W^\dagger_\mu W^\mu
                 - \onehalf m_Z^2(\hat \theta_H) Z_\mu Z^\mu 
                 - \sum_{a,b} m^F_{ab}(\hat \theta_H) \psibar_a  \psi_b ~. 
\label{effective1}
\eneq

$V_\eff (\hat \theta_H)$ is the effective potential which arises
at the one loop level.  It is finite and independent of the cutoff.\cite{YH1} 
From the second derivative at the global minimum a finite Higgs mass $m_H$
is obtained.  The finiteness of $m_H^2$  
gives a solution to the gauge-hierarchy problem. \cite{Lim1}

The mass functions $m_W(\hat \theta_H)$,  $m_Z(\hat \theta_H)$
and $m^F_{ab}(\hat \theta_H)$ arise at the tree level. 
It is essential  to include  contributions coming from
KK excited states in intermedium states.  The effective  Lagrangian
(\ref{effective1}) is obtained after integrating out all heavy KK modes.

In the $SO(5) \times U(1)$ model under consideration these mass
functions are found to be, to good accuracy in the warped space, 
\cite{SH1, HS2, HK}
\beqn
&&\hbox{\bf gauge-Higgs} \hskip 2.2cm  {\bf \big[~SM~ \big]} \cr
\noalign{\kern 10pt}
m_W(\hat \theta_H) &\sim ~~&
\frac{1}{2} g f_H \sin \hat \theta_H ~~, 
\hskip 2.0cm 
\bigg[ ~ \frac{1}{2}  g (v + H) ~ \bigg]  ~,  \cr
\noalign{\kern 10pt}
m_Z(\hat \theta_H) ~&\sim ~~&
\frac{1}{2 \cos \theta_W} g f_H \sin \hat \theta_H ~~, 
\hskip .9cm 
\bigg[ ~ \frac{1}{2 \cos \theta_W}  g (v + H) ~ \bigg] ~,   \cr
\noalign{\kern 10pt}
m_{ab}^F(\hat \theta_H) &\sim ~~& 
y_{ab}^F  f_H \sin \hat \theta_H ~~, 
\hskip 2.0cm
\Big[ ~ y_{ab}^F  (v + H) ~ \Big]   ~.
\label{effective2}
\eeqn
We have listed the formulas in the standard model in brackets on the right.
It is seen that $v+H$ in the standard model is replaced approximately
by $f_H \sin \hat \theta_H$  in the gauge-Higgs unification.  
In the standard model the mass functions are linear in the Higgs field $H$,
whereas they become periodic, nonlinear functions of $H$ in the gauge-Higgs
unification.  It is a consequence of the phase nature of $\theta_H$.
In other words the gauge invariance in the warped space naturally leads
to the nonlinear behavior.

The masses of the $W$, $Z$ bosons and fermions are given by
$m_W = \onehalf g f_H \sin \theta_H$, 
$m_Z= m_W/\cos \theta_W$,  and 
$m_{ab}^F = y_{ab}^F f_H \sin \theta_H$.  
An immediate consequence is that the Higgs couplings to $W$, $Z$, and 
fermions deviate from those in the standard model.  
In particular, 
\beqn
\begin{pmatrix} WWH \cr ZZH \cr \hbox{Yukawa} \end{pmatrix}
&=& \hbox{SM} \times \cos \theta_H ~~, \cr
\noalign{\kern 5pt}
\begin{pmatrix} WWHH \cr ZZHH  \end{pmatrix}
&=& \hbox{SM} \times \cos 2 \theta_H ~~.
\label{coupling2}
\eeqn
As we shall see below, the effective potential is 
minimized at $\theta_H= \onehalf \pi$ so that
the $WWZ$, $ZZH$ and Yukawa couplings vanish.

\mysection{5. Dynamical EW symmetry breaking}

The value of $\theta_H$ in the vacuum is determined by the location of
the global minimum of the effective potential $V_\eff (\theta_H)$, 
which  is given at the one loop level  by
\beeq
V_\eff (\theta_H) = \sum_{\rm particles}
 \pm \frac{1}{2} \int \frac{d^4 p}{(2\pi)^4} 
\sum_n \ln \Big( p^2 + m_n(\theta_H)^2  \Big) 
\label{effV1}
\eneq
where $\{ m_n(\theta_H) \}$ is a 4D mass spectrum in each KK tower
with the AB phase $\theta_H$.
As originally shown in ref.\ 1 \ignore{\bf ref.\ \cite{YH1}} 
the $\theta_H$-dependent part of 
$V_\eff (\theta_H)$ is finite, independent of how the theory is regularized.
Evaluation of $V_\eff (\theta_H)$ in the RS warped space was initiated
by Oda and Weiler.\cite{Oda1}
Since then a powerful method for evaluation has 
been developed by Falkowski.\cite{Falkowski1}

\begin{figure}[htb]
\begin{center}
%\vskip -.5cm 
\includegraphics[height=5.5cm]{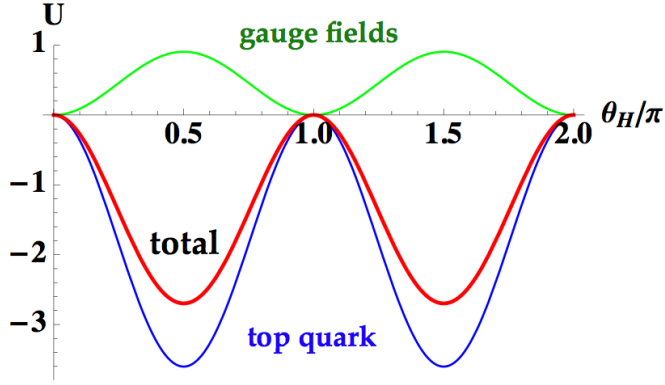}
%\vskip -.5cm
\caption{Effective potential $V_\eff (\theta_H) = m_\KK^4 /(16 \pi^4) U (\theta_H)$  
with $z_L = 10^{10}$.
\label{potential}}
\end{center}
\end{figure}

In the model under consideration the electroweak symmetry
$SU(2)_L \times U(1)_Y$ remains unbroken for $\theta_H = 0, \pi$.
Otherwise the symmetry is broken to $U(1)_\EM$. 
If there were no fermions,  the symmetry is unbroken.
Among the fermions, multiplets in the bulk containing the top quark
gives a dominant contribution to $V_\eff (\theta_H)$.  
The fact that the top quark mass (172 GeV) is larger than $m_W$
is relevant.  
Contributions from other fermions, whose masses are much smaller than
$m_W$, are negligible in the warped space.   
Contributions fom charm quarks, for instance, are suppressed by a factor $10^{-5}$.
The existence of the top 
quark triggers the electroweak symmetry breaking.  
$U(\theta_H) = (16 \pi^4/ m_\KK^4) \, V_\eff (\theta_H)$
is plotted with the warp factor $z_L = 10^{10}$ in fig.\ \ref{potential}.

It is seen that the effective potential is minimized at $\theta_H = \pm \onehalf \pi$
so that the electroweak symmetry is dynamically broken.  It follows from 
(\ref{effective2}) that $m_W = \onehalf g f_H$ and $f_H \sim 256\,$GeV. 

As the warp factor $z_L = e^{kL}$ is decreased, there appear
two kinds of phase transitions.
For $z_L < z_{L1} \sim 2200$ the  top quark mass $m_t \sim 171\,$GeV
cannot be achieved.  It necessarily becomes smaller than the observed value.
One can set the bulk mass parameter to be zero, and further decrease the 
value of $z_L$.  Then, for $z_L \sim z_{L2} \sim 1.67$ there appears 
a weakly-first-order phase transition.   For  $z_L < z_{L2}$ the global minima 
move to  $\theta_H = 0$ and $\pi$ so that the symmetry is unbroken.\footnote{There
was a numerical error in the evaluation of $V_\eff(\theta_H)$ 
in ref.\ 13.  \ignore{\bf ref. \cite{HOOS}}.}   

The curvature of the $V_\eff$ at the minimum is related to the Higgs mass
$m_H$ by 
\beeq
m_H^2 =  \frac{\pi^2 g^2 kL}{4 \,  m_\KK^2}  \, 
\frac{d^2 V_\eff}{d \theta_H^2} \bigg|_{\rm min}
~~,~~ m_\KK = \pi k z_L^{-1} ~~.
\eneq
For $z_L = 10^{10}$,  $m_\KK = 1.2\,$TeV and 
$m_H = 108\,$GeV.
At first sight one might think that this value for $m_H$ contradicts with
the LEP2 bound $m_H > 114\,$GeV.  However, the effective potential 
is minimized at $\theta_H = \pm \onehalf \pi$ where the $ZZH$ coupling
vanishes.  See (\ref{coupling2}).  The LEP2 bound is evaded as
the process $e^+ e^- \go Z^* \go ZH$ does not take place in the present model.  

\mysection{6. The absolute stability of the Higgs bosons}

In the present gauge-Higgs unification model the AB phase is dynamically 
chosen to be $\theta_H = \onehalf \pi$ at the one loop level.  
It implies that all $ZZH$, $WWH$ and
Yukawa couplings of quarks and leptons vanish so that the Higgs
boson cannot decay.  Can the Higgs boson decay  at higher orders?  
We show that the Higgs boson becomes absolutely stable
in a class of the gauge-Higgs unification models  including the 
present model.\cite{HKT}

\vskip 10pt

\leftline{\bf (i) Mirror reflection symmetry}

The theory is invariant under the mirror reflection in the fifth dimension;
\beqn
(x^\mu, y) &\go& ({x^\mu}', y')= (x^\mu, -y) ~,  \cr
\noalign{\kern 5pt}
A_M (x,y)  &\go& A_M' (x',y') = (A_\mu, -A_y)(x,y) ~,  \cr
\noalign{\kern 5pt}
\Psi_a (x,y)  &\go&  \Psi_a' (x',y') = \pm \gamma^5 \Psi_a (x,y) ~.
\label{reflect1}
\eeqn
It implies that physics is invariant under
\beeq
\hat \theta_H (x) = \theta_H + \frac{H(x)}{f_H} \go 
\hat \theta_H' (x') = - \hat \theta_H(x)
\label{reflect2}
\eneq
while all other SM particles remain unchanged.

\vskip 10pt

\leftline{\bf (ii) Enhanced gauge invariance}

One may notice that the effective potential $V_\eff (\theta_H)$ at the 
one loop level depicted in fig.\ \ref{potential} has periodicity $\pi$.
Indeed, this remains true to all order.  Gauge
invariance is enhanced. 

Consider a lare gauge transformation  (\ref{largeGT1}) with $\alpha=\pi$.
With $\Omega^{\rm large}(y; \pi)$ the boundary conditions change to
$({P_0'}^{\rm vec}, {P_1'}^{\rm vec} ) = ({P_0}^{\rm vec}, {P_1}^{\rm vec})$
in the vectorial representation and 
$({P_0'}^{\rm sp}, {P_1'}^{\rm sp} ) = ({P_0}^{\rm sp}, -{P_1}^{\rm sp})$
in the spinorial representation.
The AB phase changes to $\theta_H' =  \theta_H + \pi$.
In the present model all bulk fermions are in the vector representation.
The brane fermions and scalar field on the Planck brane are not affected by 
this transformation as  $\Omega^{\rm large}(0; \pi) = 1$.
Hence the theory is invariant under $\theta_H \go \theta_H + \pi$.
All physical quantities become periodic in $\theta_H$ with a reduced period $\pi$.

\vskip 10pt

\leftline{\bf (iii) H-parity}

In a class of the  $SO(5) \times U(1)$ gauge-Higgs unification models
in the warped space which contains bulk fermions only in tensorial representations
of $SO(5)$ and brane fermions only on the Planck brane, 
the enhanced gauge symmetry with the mirror reflection symmetry  leads to
\beqn
&&\hskip -1cm
V_\eff (\hat \theta_H + \pi) = V_\eff (\hat \theta_H) 
 = V_\eff (- \hat \theta_H) ~, \cr
\noalign{\kern 5pt}
&&\hskip -1cm
m_{W, Z}^2(\hat \theta_H + \pi) = m_{W, Z}^2 (\hat \theta_H) 
= m_{W, Z}^2 (- \hat \theta_H)~, \cr
\noalign{\kern 5pt}
&&\hskip -1cm
m^F_{ab}(\hat \theta_H + \pi) = - m^F_{ab} (\hat \theta_H) 
= m^F_{ab}  (- \hat \theta_H) ~.
\label{symmetry1}
\eeqn
$m_{W, Z}(0) = m^F_{ab}  (0) = 0$  as the EW symmetry is recovered 
at $\theta_H=0$.

In the previous section we have seen that $V_\eff(\theta_H)$ is 
minimized at $\theta_H = \onehalf \pi$.  It follows then from (\ref{symmetry1})
that all of $V_\eff (\hat \theta_H)$, $m_{W, Z}^2 (\hat \theta_H)$ and
$m^F_{ab} (\hat \theta_H)$ satisfy a relation 
$F(\onehalf \pi + f_H^{-1} H) = F(\onehalf \pi - f_H^{-1} H)$.  
They are even functions of $H$ when expanded around $\theta_H = \pm \onehalf \pi$.
All odd-power Higgs couplings  $H^{2\ell +1}$,  
$H^{2\ell +1} W_\mu^\dagger W^\mu$, $H^{2\ell +1} Z_\mu  Z^\mu$, 
and $H^{2\ell +1} \psibar_a \psi_b$,  vanish.

The effective interactions at low energies are invariant
under $H(x) \go - H(x)$ with all other fields kept intact
at $\theta_H = \pm \onehalf \pi$.   There arises  the $H$-parity invariance.   
Among low energy fields only the Higgs field is $H$-parity odd.
The Higgs boson becomes absolutely stable, protected by the $H$-parity conservation.

\mysection{7. Stable Higgs bosons as cold dark matter}

Abosolutely stable Higgs bosons become cold dark matter (CDM) 
in the present universe.  
They are copiously produced in the very early universe.  
As the annihilation rate of Higgs bosons falls below the expansion rate 
of the universe,  the annihilation processes get effectively  frozen
and the remnant Higgs bosons become dark matter.\cite{HKT} 

The annihilation rates can be estimated from the effective Lagrangian
(\ref{effective1}) with (\ref{effective2}).   At $\theta_H = \onehalf \pi$ one has
\beeq
{\cal L}_\eff \sim - \Big\{ m_W^2 W_\mu^\dagger W^\mu 
     + \frac{1}{2} m_Z^2 Z_\mu Z^\mu \Big\}  \cos^2 \frac{H}{f_H} 
- \sum_a m_a \psibar_a \psi_a \cos \frac{H}{f_H} ~.
\label{effective3}
\eneq
As $m_W \sim \onehalf g f_H$,   $f_H$ is determined to be $ \sim 246 \,$GeV. 
Although the Yukawa coupling $\psibar_a \psi_a H$ vanishes, 
the $\psibar_a \psi_a H^2$ coupling is nonvanishing,   given by
$(m_a/2 f_H^2) \psibar_a \psi_a H^2$.  It is generated by two vertices 
$\psi_a \psi_a^{(n)} H$ where $\psi_a^{(n)} $ is the $n$-th
KK excited state of $\psi_a$. 

The Higgs mass $m_H$  is predicted in the present model in the range of
50 GeV to 130 GeV, depending on the value of the warp factor.
If $m_H > m_W$, the dominant annihilation modes are  $HH \go WW, \, ZZ$.
The rate is large so that the resultant relic abundance becomes very small.
For $m_H < m_W$ the relevant annihilation modes are 
$HH \go W^* W^*, \, Z^* Z^*$, $b \bar b, c \bar c, \tau \bar \tau$, and
$ gg$.  Here $W^*$ ($Z^*$) indicates virtual $W$ ($Z$) which subsequently
decays into a fermion pair.  Annihilation into a gluon ($g$) pair takes place 
through a top quark  loop.

\begin{figure}[hbt]
\begin{center}
\includegraphics[height=8.cm]{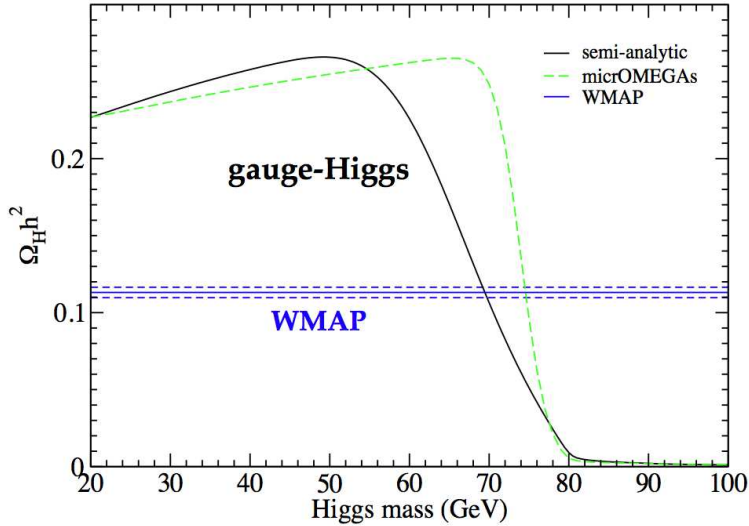} 
\caption{{\bf }  Thermal relic density of Higgs boson DM with $f_H = 246\,$GeV.
The solid curve is obtained by the semi-analytic formulae.
The horizontal band is the WMAP data 
$\Omega_{\rm CDM} h^2 = 0.1131 \pm 0.0034$.
The value in the WMAP data is obtained with $m_H \sim 70\,$GeV.
}
\label{relic1}
\end{center}
\end{figure}

The relic abundance of  Higgs bosons is evaluated as a function of 
$m_H$.  It is depicted in fig.\  \ref{relic1}.
It is seen that  $\Omega_H h^2$ determined from the WMAP data
is reproduced in the gauge-Higgs unification with $m_H \sim 70\,$GeV.
It is remarkable that the gauge-Higgs unification scenario gives
$\Omega_H h^2$ in the just right order of magnitude for 
$20 \,{\rm GeV} < m_H < 75 \,{\rm GeV}$.  
With $m_H = 70\,$GeV, the freeze-out temperature is $T_f \sim 3\,$GeV.
The relative contributions of the $HH \go W^* W^*$ and $b \bar b$ modes
are 61\% and 34\%, respectively.

If Higgs bosons constitute the cold dark matter of the universe, 
they can be detected by observing Higgs-nucleon elastic scattering 
process, $HN \go HN$.
The relevant part of the effective interaction
(\ref{effective3}) is
${\cal L}_\eff=   (H^2/ 2 f_H^2) \sum_f m_f  \bar f f$.
With QCD corrections taken into account the effective interaction for
the Higgs-nucleon coupling becomes
\beeq
{\cal L}_{HN} \simeq \frac{2+7f_N}{9} ~\frac{m_N}{2 f_H^2} ~H^2
\overline{N} N ~ , 
\label{effective4}
\eneq
where $f_N=\sum_{q=u,d,s}f^N_q$ and 
$\langle N|m_q\bar q q|N\rangle=m_N f^N_q$.
With this coupling (\ref{effective4})  the  spin-independent (SI)
Higgs-nucleon scattering cross section is found to be
\beeq
\sigma_{\rm SI} \simeq
\frac{1}{4\pi} \bigg( \frac{2 + 7 f_N}{9} \bigg)^2 
\frac{m_N^4}{f_H^4 ( m_H + m_N)^2} ~,
\label{SIrate}
\eneq
in the non-relativistic limit.
There is ambiguity in the value of $f_N$.  
Our prediction is $\sigma_{SI}\simeq (1.2 - 2.7)\times 10^{-43}\,\mathrm{cm}^2$
for $f_N = (0.1 - 0.3)$.  
The present experimental upper bounds for
the spin-independent WIMP-nucleon cross sections come from 
CDMS II\cite{Ahmed:2008eu} and XENON10 \cite{Angle:2007uj}.
From the recent CDMS II data
$\sigma_\mathrm{SI} \lesssim 7 \times 10^{-44}\,\mathrm{cm}^2$ 
at 90 \% CL with the WIMP mass $70\,$GeV.
With many uncertainties and ambiguity in the analysis taken into account, 
this does not necessarily mean that the present  gauge-Higgs unification model
is excluded.   The next generation experiments for the direct detection of 
WIMP-nucleon scattering are awaited to pin down the rate.

\mysection{8. Stable Higgs bosons at Tevatron/LHC/ILC}

In all of the experiments performed so far, Higgs bosons are
searched by trying to  identify their decay products.  If Higgs bosons are stable,
however, this way of doing experiments becomes a vain effort. 

Higgs bosons are produced in pairs.  Typical processes are
$Z^* \go ZHH$, $W^* \go WHH$, $WW \go HH$, $ZZ \go HH$, and $gg \go HH$.
Higgs bosons are stable so that they appear as missing energies and momenta
in collider experiments.  The appearance of two particles of missing energies 
and momenta in the final state makes experiments hard, but not impossible.  
These events must be distinguished from those involving neutrinos.
Since $m_H \sim  70\,$GeV,  Higgs bosons can be copiously produced
at LHC.  It is a challenging task to identify stable Higgs bosons at colliders.

The effects of the vanishing $WWH$ and $ZZH$ couplings can be
seen in $WW$, $WZ$, and $ZZ$ elastic scattering, too.  These scattering
amplitudes become large as the energy is increased, much faster than 
in the standard model.\cite{sakamura2} 

\mysection{9. Gauge couplings of quarks and leptons}

We have seen large deviations in the Higgs couplings from the standard model.
There arise deviations in the gauge couplings of quarks and leptons as well,
which have to be scrutinized with the electroweak precision 
measuremens.\cite{HNU, Agashe1, HNSS}

5D gauge couplings of fermions are universal.  However, 4D gauge couplings 
of quarks and leptons are obtained by inserting their wave functions and $W$ ($Z$)
wave function into 5D Lagrangian and integrating over the fifth coordinate.
Since the wave functions depend on quarks and leptons, slight variations
appear in 4D gauge couplings, resulting in violation of the universality.
The wave functions of $W$ and $Z$ bosons are determined in
refs.\ \cite{SH1} and \cite{HS2},  
whereas those of quarks and leptons are determined in 
refs.\ \cite{HNU} and \cite{HK}. \ignore{\bf refs.\ \cite{HK, HNU}. }

There arise deviations in the $W$ boson couplings from the standard model.
For the $t$ and $b$ quarks their interactions are given by 
\beeq
{1\over 2} g^{(W)}_{tb,L} 
  (W_\mu  \bar{b}_L \gamma^\mu t_L 
  +W_\mu^\dag  \bar{t}_L \gamma^\mu b_L) 
 +  {1\over 2} g^{(W)}_{tb, R}
  (W_\mu  \bar{b}_R \gamma^\mu t_R 
  +W_\mu^\dag  \bar{t}_R \gamma^\mu b_R).
\label{Wcoupling1}
\eneq
Not only left-handed components but also right-handed  components
couple to $W$ in general.  The $W$ coupling defined experimentally is 
the coupling between left-handed $e$ and $\nu_e$, $g^{(W)}_{e\nu,L} $.
In the standard model the couplings are universal; 
$g^{(W)}_{f,L} = g^{(W)}_{e\nu,L} $ and  $g^{(W)}_{f,R} = 0$.  
The violation of the universality for the left-handed quarks and leptons, 
$g^{(W)}_{f,L} /g^{(W)}_{e\nu,L} -1$,  is summarized in Table 1.
The $W$ couplings to the right-handed quarks and leptons are summarized in Table 2.
One can see that the deviation from the standard model is  extremely tiny except for 
top-bottom,   well below the current limits.
For the left-handed top-bottom quarks the deviation is about 2\%.

\begin{table}[b]
%\vskip 10pt
%\noalign{\kern 5pt}
%\tbl{Comparison of acoustic for frequencies for piston-cylinder problem.}
\begin{center}
\caption{The deviation of the $W$  couplings for left-handed 
leptons and quarks from the standard model for the warp factor $z_L = 10^{15}$.}
\begin{tabular}{|ccccc|}
%\hline
\noalign{\kern 10pt}
\multicolumn{5}{c}{$g^{(W)}_{fL} /g^{(W)}_{e\nu,L} -1$} \\ 
\noalign{\kern 3pt}
\hline
 $\nu_{\mu},\mu$  & $\nu_{\tau},\tau$ & $u,d$ & $c,s$  & $t,b$  \\ \hline
$-1.022\times\!\! 10^{-8}$  & %-1.02237*10^-8
$-2.736\times\!\! 10^{-6}$   & %-2.73558*10^-6
$-2.778\times\!\! 10^{-11}$ & %-2.77842*10^-11
$-1.415\times\!\! 10^{-6}$  & %-1.41482*10^-6
$-0.02329$ %-0.0232888
 \\ \hline
\end{tabular}
\end{center}
\end{table}

\begin{table}[t]
\vskip 10pt
%\noalign{\kern 5pt}
%\tbl{Comparison of acoustic for frequencies for piston-cylinder problem.}
\begin{center}
\caption{The  $W$  couplings for right-handed  leptons and quarks for the warp factor $z_L = 10^{15}$.  They are all tiny.}
\begin{tabular}{|ccc|}
%\hline
\multicolumn{3}{c}{$g^{(W)}_{fR} /g^{(W)}_{e\nu,L}$} \\ 
\noalign{\kern 5pt}
\hline
 $\nu_{e}, e$  &  $\nu_{\mu},\mu$  & $\nu_{\tau},\tau$   \\ \hline
 $-4.761\times\!\! 10^{-21}$ & %-4.76119*10^-21
$-4.141\times\!\! 10^{-16}$ & %-4.14068*10^-16
$-1.133\times\!\! 10^{-13}$   %-1.13266*10^-13
\\ \hline
\noalign{\kern 3pt} \hline
$u,d$ & $c,s$  & $t,b$ \\ \hline
 $-1.392\times\!\! 10^{-11}$ & %-1.39187*10^-11
$-1.896\times\!\! 10^{-7}$  & %-1.89557*10^-7
$-0.001264$        
 \\ \hline
\end{tabular}
\end{center}
\end{table}
\begin{table}[h]
%\vskip 10pt
%\noalign{\kern 5pt}
%\tbl{Comparison of acoustic for frequencies for piston-cylinder problem.}
\begin{center}
\caption{The deviation of the $Z$  couplings for  
leptons and quarks from the standard model for the warp factor $z_L = 10^{15}$. 
In the column SM  the values in the standard model are quoted.}
\begin{tabular}{|c|ccc|c|}
%\hline
\multicolumn{5}{c}{$\tilde g^{(Z)}_{f,LR} = g^{(Z)}_{f,LR} / g^{(W)}_{e\nu,L}$} \\ 
%\hline
\noalign{\kern 5pt}
 \hline
$f $& $\nu_e$ &  $\nu_{\mu}$  & $\nu_{\tau}$ & SM   \\ \hline
$\tilde g^{(Z)}_{f,L} $  & 
0.500822 & %0.500822
0.500822 & %0.500822
0.500822 & %0.500822
0.5
 \\ \hline
 $\tilde g^{(Z)}_{fR}$
 & 
 $-5.754\times\!\! 10^{-31}$ & %-5.75391*10^-31
$-5.392\times\!\! 10^{-29}$ & %-5.39201*10^-29
$-1.840\times\!\! 10^{-27}$ & %-1.84021*10^-27
 0
\\ \hline
\noalign{\kern 4pt}
\hline
& $e$ & $\mu$ & $\tau$ &   \\ \hline
$\tilde g^{(Z)}_{fL}$  & 
$-0.2692$ & %-0.269242
$-0.2692$ & %-0.269242
$-0.2692$ & %-0.269239
$-0.2688$   %-0.2688
\\ \hline
$\tilde g^{(Z)}_{fR}$
 & 
0.2334 & %0.233393
0.2333 & %0.233335
0.2333 & %0.2333
0.2312   %0.2312
\\ \hline
\noalign{\kern 4pt}
\hline
& $u$ & $c$ & $t$ &   \\ \hline
$\tilde g^{(Z)}_{fL}$  &
0.3464 & %0.346435
0.3464 & %0.346434
0.3204 & %0.320407
0.3459   %0.345867
\\ \hline
$\tilde g^{(Z)}_{fR}$
& 
$-0.1556$ & %-0.155579
$-0.1556$ & %-0.155536
$-0.1823$ & %-0.182329
$-0.1541$   %-0.154133
\\ \hline
\noalign{\kern 4pt}
\hline
& $d$ & $s$ & $b$ &   \\ \hline
$\tilde g^{(Z)}_{fL}$  &
$-0.4236$ & %-0.423628
$-0.4236$ & %-0.423628
$-0.4241$ & %-0.424107
$-0.4229$   %-0.422933
\\ \hline
$\tilde g^{(Z)}_{fR}$
& 
0.07779 & %0.0777897
0.07777 & %0.0777674
0.07774 & %0.0777438
0.07707   %0.0770667
 \\ \hline 
\end{tabular}
\end{center}
\end{table}
%

%\eject

Similarly the $Z$ boson couplings deviate from the standard model.
The  couplings of $t$ and $b$ quarks take the form
\beeq
\frac{1}{\cos\theta_W} Z_\mu \bigg\{ 
g^{(Z)}_{tL}   \bar{t}_L \gamma^\mu t_L
+g^{(Z)}_{tR} \bar{t}_R \gamma^\mu t_R  
+g^{(Z)}_{bL} \bar{b}_L \gamma^\mu b_L
+g^{(Z)}_{bR} \bar{b}_R \gamma^\mu b_R   \bigg\}  ~.
\label{Zcoupling1}
\eneq
The relevant couplings are $g^{(Z)}_{f,LR} / g^{(W)}_{e\nu,L}$. 
They  are summarized  in Table 3, in which the couplings in the 
standard model are also listed for comparison.

It is seen  from the table, the deviation from the standard model is rather small
(0.1\% to 1\%)  except for the top quark coupling for $z_L = 10^{15}$.  
The $Z$ couplings of  the left- and right-handed top quark deviate from those in 
the standard model by $-7$\% and 18\%, respectively.
The deviations in the $Z  b_L \bar b_L $ and $Z b_R \bar b_R$ couplings 
are 0.3\% and 0.9\%, respectively.

\begin{table}[t]
\vskip 5pt
%\noalign{\kern 5pt}
\begin{center}
\caption{The forward-backward asymmetry on the $Z$ pole in $e^+ e^-$ collisions.
The numbers in the gauge-Higgs unification scenario with the warp factor 
$z_L = 10^{15}$ are quoted from Uekusa, ref.\ 24.}
\begin{tabular}{|c|c|c|c|}
%\hline
\multicolumn{4}{c}{$A_{FB}^f$} \\ 
%\hline
\noalign{\kern 3pt}
\hline
~$f $~ & Experiment &  ~Gauge-Higgs~  & Standard Model   \\ \hline  %\hline
$c $  & 
~$0.0707 \pm 0.0035$~ & 
0.07073 & 
$0.0738 \pm 0.0006$
 \\ \hline
$s$ & 
$0.0976 \pm 0.0114$ & 
0.09950 & 
$0.1034 \pm 0.0007$
\\ \hline
$b$  & 
$0.0992 \pm 0.0016$ & 
0.09952 & 
$0.1033 \pm 0.0007$
\\ \hline
$e$ & 
$0.0145 \pm 0.0025$ & 
0.01511 & 
$0.01627 \pm 0.00023$
\\ \hline
$\mu$ & 
$0.0169 \pm 0.0013$ & 
0.01513 & 
\\ \hline
$\tau$ & 
$0.0188 \pm 0.0017$ & 
0.01515 & 
\\ \hline
\end{tabular}
\end{center}
\end{table}

There results an important prediction for the forward-backward asymmetry
in the $e^+ e^-$ collisions on the $Z$ pole as pointed out by Uekusa.\cite{uekusa}
The asymmetry is given by
\beeq
A_{FB}^f = \frac{3}{4} A_{LR}^e A_{LR}^f ~~,~~
A_{LR}^f  = \myfrac{\big(g_{fL}^{(Z)} \big)^2 - \big(g_{fR}^{(Z)} \big)^2}
{\big(g_{fL}^{(Z)} \big)^2 + \big(g_{fR}^{(Z)} \big)^2} ~.
\label{FBasym}
\eneq
The predictions are summarized in Table 4.
It is exciting that the  gauge-Higgs unification model gives better fits
to the observed data for the asymmetry in $b \bar b$ and $c \bar c$ 
than the standard model.    CP violation,   anomalous magnetic moment, 
and  electric dipole moment in gauge-Higgs unification have been also discussed.\cite{Lim4}

\eject

\mysection{10. Summary}

The $SO(5) \times U(1)$ gauge-Higgs unification in the Randall-Sundrum warped
spacetime is promising.  The Higgs boson becomes a part of the gauge fields in
higher dimensions, and is unified with gauge bosons.  
The resultant 4D gauge couplings of quarks and leptons
are close to those in the standard model.  The Higgs couplings, on the other hand,
deviate significantly from those in the standard model.

In a large class of the  $SO(5) \times U(1)$ gauge-Higgs unification models
the Higgs boson becomes absolutely stable to all order in perturbation 
theory.  In the evolution of the universe Higgs bosons become  cold dark matter.
From the WMAP data the Higgs boson mass is determined to be around 70$\,$GeV.

In collider experiments Higgs bosons are produced in pairs.  They appear as
missing energies and momenta.  The way of searching for Higgs bosons must be
altered.

\vskip 20pt

\leftline{\bf Acknowledgment}
\vskip 5pt

This work was supported in part 
by  Scientific Grants from the Ministry of Education and Science, 
Grant No.\ 20244028, Grant No.\ 20025004,  and Grant No.\ 50324744.

\vskip 20pt 

\leftline{\bf References}

\renewenvironment{thebibliography}[1]
         {\begin{list}{[$\,$\arabic{enumi}$\,$]}  % {\arabic{enumi}.}
         {\usecounter{enumi}\setlength{\parsep}{0pt}
          \setlength{\itemsep}{0pt}  \renewcommand{\baselinestretch}{1.2}
          \settowidth
         {\labelwidth}{#1 ~ ~}\sloppy}}{\end{list}}

\end{document}